\documentclass[aps,prc,reprint,superscriptaddress,showpacs,showkeys,amsmath,amssymb]{revtex4-1}
\usepackage{graphicx}
\usepackage{float}
\usepackage{dcolumn}
\usepackage{bm}
\usepackage{lineno}
\usepackage{rotating}
\usepackage{url}
\usepackage{flushend,cuted}
\usepackage[colorlinks,citecolor=blue,linkcolor=blue,hypertex,breaklinks=true]{hyperref}
\begin{document}

\title{Structural Effects and Competition Mechanisms Targeting the Interactions between p53 and Mdm2 for Cancer Therapy}

\author{Shuxia Liu}
\affiliation{College of Nuclear Science and Technology,
Beijing Normal University, Beijing 100875, China}

\author{Yizhao Geng}
\affiliation{School of Science, Hebei University of Technology, Tianjin 300401, China}

\author{Shiwei Yan}
\email[Email address:]{yansw@bnu.edu.cn}
\affiliation{College of Nuclear Science and Technology,
Beijing Normal University, Beijing 100875, China}
\affiliation{Beijing Radiation Center, Beijing 100875, China}

\date{\today}

\begin{abstract}
About half of human cancers show normal TP53 gene and aberrant overexpression of Mdm2 and/or MdmX. This fact promotes a promising cancer therapeutic strategy which targeting the interactions between p53 and Mdm2/MdmX. For developing the inhibitors to disrupt the p53-Mdm2/MdmX interactions, we systematically investigate structural and interaction characteristics of p53 and inhibitors with Mdm2 and MdmX from atomistic level by exploiting stochastic molecular dynamics simulations. We find that some specific $\alpha$ helices in Mdm2 and MdmX structure play key role in their bindings with inhibitors and the hydrogen bond formed by  residue Trp23 of p53 with its counterpart in Mdm2/MdmX determines dynamical competition processes of the disruption of Mdm2-p53 interaction and replacement of p53 from Mdm2-p53 complex {\it in vivo}. We hope that the results reported in this paper provide basic information for designing functional inhibitors and realizing cancer gene therapy.
\end{abstract}

\maketitle

\section{Introduction\label{sect-1}}
p53, a tumor suppressor protein, is the guardian of the genome and plays a crucial role in the regulation of cell cycle, apoptosis, DNA repair and angiogenesis \cite{ACIE-50-2680}. p53 concentration increase and, thus, cause cell cycle arrest or apoptosis response to the signal of DNA damage. The activation of p53 can also activate another kind of protein, Mdm2, which negatively regulate the concentration of p53. In normal cells, both concentrations of Mdm2 and p53 are low that provide growth advantage to cells. However, about half of human cancers show normal TP53 gene and aberrant overexpression of Mdm2 and/or MdmX \cite{JBC-286-23725} that inhibit the activation of p53. This fact provides a potent strategy for cancer therapy, i.e. restoration the activity of p53 by inhibitors which can occupy the p53-binding site of Mdm2 and inhibit the interaction of p53 and Mdm2. Once freed from Mdm2, p53 rapidly accumulates in the nuclei of cancer cells, activates p53 target genes and the p53 pathway, resulting in cell-cycle arrest and apoptosis \cite{Science-303-844,Science-274-948,NRC-3-102}.

In history, it has been difficult to develop small-molecule inhibitors of nonenzyme protein-protein interactions \cite{Science-303-844}. Protein-protein interactions usually involve large and flat surfaces that are difficult to break by low molecular weight compounds \cite{NRDD-3-301,JMM-83-955}. However,
the crystal structure of Mdm2 bound to a peptide from the transactivation domain of p53 has revealed that Mdm2 possesses a relatively deep hydrophobic pocket that is filled primarily by three side chains from the helical region of the peptide \cite{Science-274-948}. The existence of such a well-defined pocket on the Mdm2 molecule raised the expectation that compounds with low molecular weights could be found that would block the interaction of Mdm2 with p53.

In terms of the characteristic of p53 binding to Mdm2, the design of inhibitors should follow the principle that the inhibitor can model the Mdm2-binding site of p53 \cite{NRC-3-102}. The interactions between p53 and Mdm2 are mainly hydrophobic interactions and hydrogen bonds. Three residues of p53, Phe19, Trp23 and Leu26, have been found essential for the binding between p53 and Mdm2 and they are inserted into a deep hydrophobic pocket on the surface of the MDM2 molecule. These features of interactions between p53 and Mdm2 should be maintained in the design of efficient inhibitor. The interface between p53 and Mdm2 is small that permits the design of relatively small inhibitors which have higher oral bioavailability \cite{Structure-21-2143}.

Recently, rational designing with molecular docking
and high throughput virtual screening approaches has led to the
generation of diversified set of small molecules and peptides that
can restore the activity of p53 \cite{NRDD-13-217,PBMB-117-250,NRC-3-102}. In these inhibitors, Nutlin family is a kind of non-polypeptide inhibitor which include Nutlin1, Nutlin2, Nutlin3, RG7112 and {\it et.~al.}~\cite{CR-73-2587}. One of these promising Mdm2 inhibitors is Nutlin3 which is currently under clinical investigations. It has been shown that Nutlin3 can tightly combine to Mdm2 and efficiently inhibit the p53-Mdm2 interaction \cite{JACS-136-18023}.

Moreover, MdmX (Mdm4 in mice) is another negative regulator of p53, control the stability and activity of p53 in a different mechanism to Mdm2 \cite{MCR-1-1001}. MdmX is highly homologous to Mdm2, with the
difference that it does not possess any ubiquitin ligase activity and does not cause p53 degradation although it binds to the N-terminus of p53 and suppresses p53 transcriptional activities \cite{EMBO-R-2-1029}. However, even Mdm2 and MdmX are homologous proteins with high degree of homolog especially in their N-terminal p53 binding domain, existing inhibitors of Mdm2-p53 have weak function on interactions of MdmX-p53 that highly reduce the effect of inhibitors in cancer therapy. Some has suggested the importance to develop dual small-molecule inhibitors of p53-Mdm2/p53-MdmX interactions for the complete reactivation of p53 \cite{ARPT-49-223}.

\begin{figure*}[hptb]
\setlength{\abovecaptionskip}{-15pt}
\begin{center}
\includegraphics[width=0.85\linewidth]{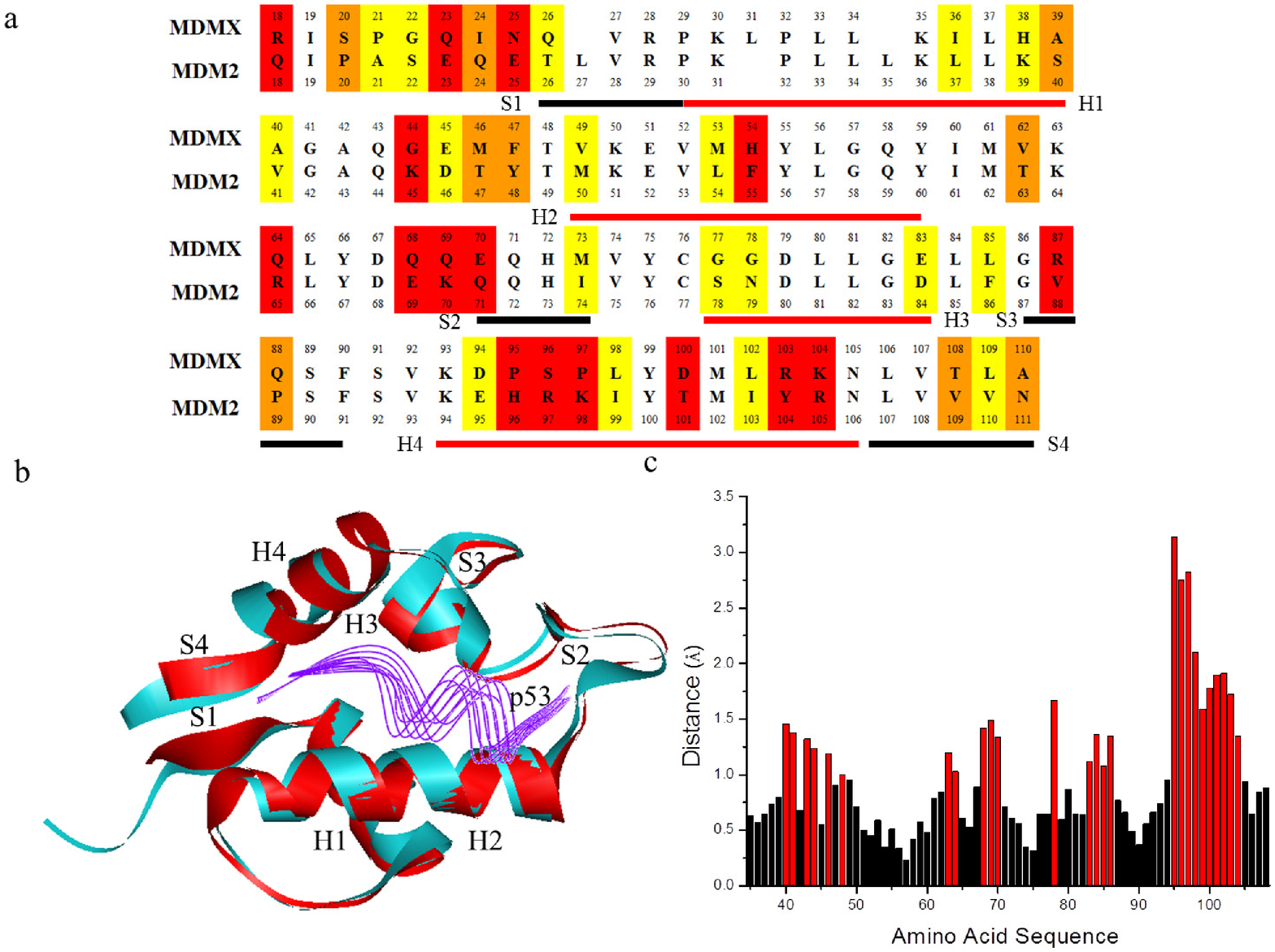}
\end{center}
\caption{Comparison of sequences and structures of Mdm2/MdmX\textquoteright s p53-binding sites. (a) Amino acid sequence alignment of p53-binding sites of Mdm2 and MdmX. The different amino acids are highlighted by different colors which represent the three types with changes of polarity(red), hydrophobicity (brown) and geometric structures (yellow). Abbreviations for the amino acid residues are: A, Ala; C, Cys; D, Asp; E, Glu; F, Phe; G, Gly; H, His; I, Ile; K, Lys; L, Leu; M, Met; N, Asn; P, Pro; Q, Gln; R, Arg; S, Ser; T, Thr; V, Val; W, Trp; Y, Tyr. (b) Superposition of crystal structures of N-termini of Mdm2 (red) and MdmX (green). p53 is depicted with purple lines. The large structural differences appear on two loop regions and two $\alpha$ helices (H3 and H4). (c) Every pairs of C$_\alpha$ distances of the structure in (b). The red region show the the correspondant pairs of C$_\alpha$ whose distances  are longer than 1 \AA. Due to the uncertainty of H1 structure of MdmX, the comparison of distance between two H1s are not shown.}\label{fig1}
\end{figure*}

To this end, the structural characteristics and interaction properties of inhibitor-Mdm2 and inhibitor-MdmX should be systematically investigated. Consider the fact that the currently used docking scoring functions are not expected to consistently provide accurate predictions of the protein-ligand binding free energies for all of the protein-ligand binding systems and are difficult to account for the effects of protein dynamics on the microscopic binding during the simple docking process \cite{JPCB-110-26365}, we will exploit molecular dynamics (MD) simulations that can more reasonably account for the solvation effects and the dynamics of the protein-ligand binding. The MD simulations
allow us to obtain a dynamically stable protein-ligand binding mode associated with a stable MD trajectory.

In this paper, we set up four models (p53-Mdm2, p53-MdmX, Nutlin3-Mdm2 and Nutlin3-MdmX) to investigate the interactions between p53-Mdm2/MdmX and between Nutlin3-Mdm2/MdmX by using MD simulations through comparison of interactions between p53-Mdm2/MdmX and between inhibitor-Mdm2/MdmX. We will find that the binding patterns of p53-Mdm2 and p53-MdmX are same despite of some differences in the sequences and structures between Mdm2 and MdmX. Contrary to the tightly binding of Nutlin-3 to Mdm2, Nutlin-3 can not bind to MdmX due to three factors arise from such sequence difference between Mdm2 and MdmX.

\section{Methods\label{sect-2}}
In the modeling, complexes of p53-Mdm2 (PDB ID: 1YCR\cite{Science-274-948}), p53-MdmX (PDB ID: 3DAB\cite{CC-7-2441}) and Nutlin3-Mdm2 (PDB ID: 4HG7\cite{ACSD-69-1358}) are extracted from the corresponding PDB files and solvated by explicit water molecules, respectively. The radius of water sphere is 10 {\AA} larger than that of proteins. Because there is still no crystal structure of Nutlin3-MdmX complex, we replace the Mdm2 molecule of 4HG7 by MdmX molecule to construct Nutlin3-MdmX complex. This complex is also buried by explicit water molecules.

The ionic concentration is 150mM. TIP3P\cite{JCP-79-926} is used to model water molecules. The protein-water complexes are performed 30,000 steps energy minimization and then performed 50 nanoseconds (ns) MD simulations. In our MD simulations, $\alpha$-carbons of ILE500, ASP67, LEU109 in MdmX and GLU25, VAL109 in Mdm2 are fixed. The models are made by VMD (version 1.9) \cite{JMG-14-33} and MD simulations are performed by NAMD code (version 2.9)\cite{JCC-26-1781} with the force file CHARMM \cite{JPC-102-3586} at constant temperature of 310 K. The non-bonded Coulomb and van der Waals interactions are calculated with a cutoff using a switching function starting at a distance of 13 {\AA} and reaching zero at 15 {\AA}. The integration time step is 2 femtoseconds (fs).

\section{Results and discussions\label{sect-3}}

\subsection{Comparison of sequences and structures between MdmX and Mdm2\label{sect-3-1}}

The functions of a bio-molecule are mainly determined by its structure. As the starting point, we investigate the sequence and structure difference between MdmX and Mdm2. Indeed, the fact that existing Mdm2 inhibitors have weaker binding affinity for MdmX protein \cite{PNAS-105-3933,JBC-281-33030,Nature-444-61} may indicate that some structural difference between p53-binding domains of Mdm2 and MdmX play crucial role in the binding properties of inhibitors to Mdm2 and MdmX \cite{JACS-136-18023,ARPT-49-223}.

\begin{figure*}[tbph]
\setlength{\abovecaptionskip}{-5pt}
\begin{center}
\includegraphics[width=0.85\linewidth]{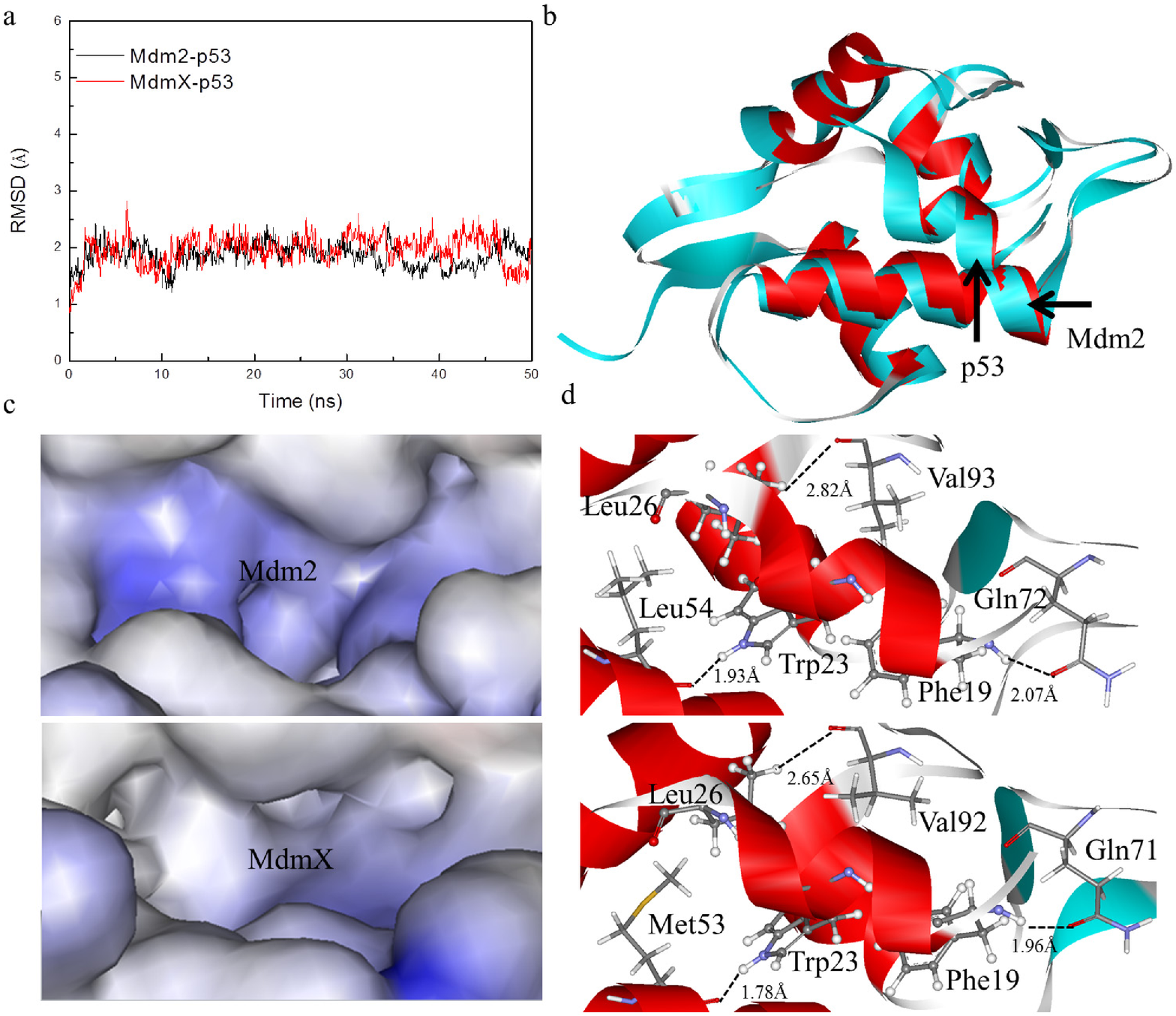}
\end{center}
\caption{Simulation results of p53-Mdm2 and p53-MdmX complexes. (a) RMSD of protein-protein distance in p53-Mdm2 and p53-MdmX complexes relative to their initial structures. After 5 ns, the two systems reach stable state. (b) Superposition of two stable p53-Mdm2 (red) and p53-MdmX (blue) complex structures. (c) The
hydrophobic surface of the p53-binding sites of Mdm2 and MdmX. The volume of this site
on Mdm2 are slightly smaller than that on MdmX. (d) The three hydrogen bonds formed between PHE19, TRP23 and LEU26 of p53 protein and GLN72/GLN71, LEU54/MET53 and
VAL93/VAL92 of Mdm2/MdmX. The distances between O and H atoms of these hydrogen
bonds are explicitly shown.}\label{fig2}
\end{figure*}

\begin{figure}[hbtp]
\setlength{\abovecaptionskip}{-5pt}
\begin{center}
\includegraphics[width=1.0\linewidth]{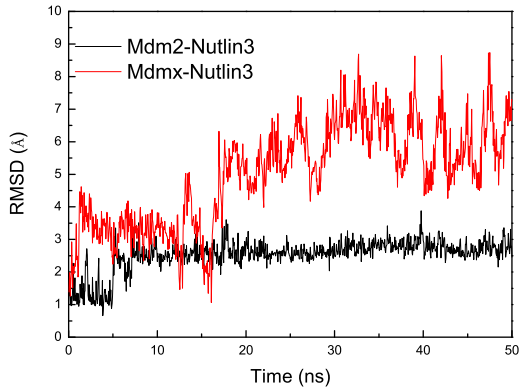}
\end{center}
\caption{RMSD of distance between Nutlin3 and Mdm2 (black), as well as  Nutlin3 and MdmX (red), respectively. Mdm2-Nutlin3 reaches stable state after 5 ns. However, MdmX-Nutlin3 can not reach a
stable structure.}\label{fig3}
\end{figure}

\begin{figure*}[htbp]
\setlength{\abovecaptionskip}{-5pt}
\begin{center}
\includegraphics[width=0.85\linewidth]{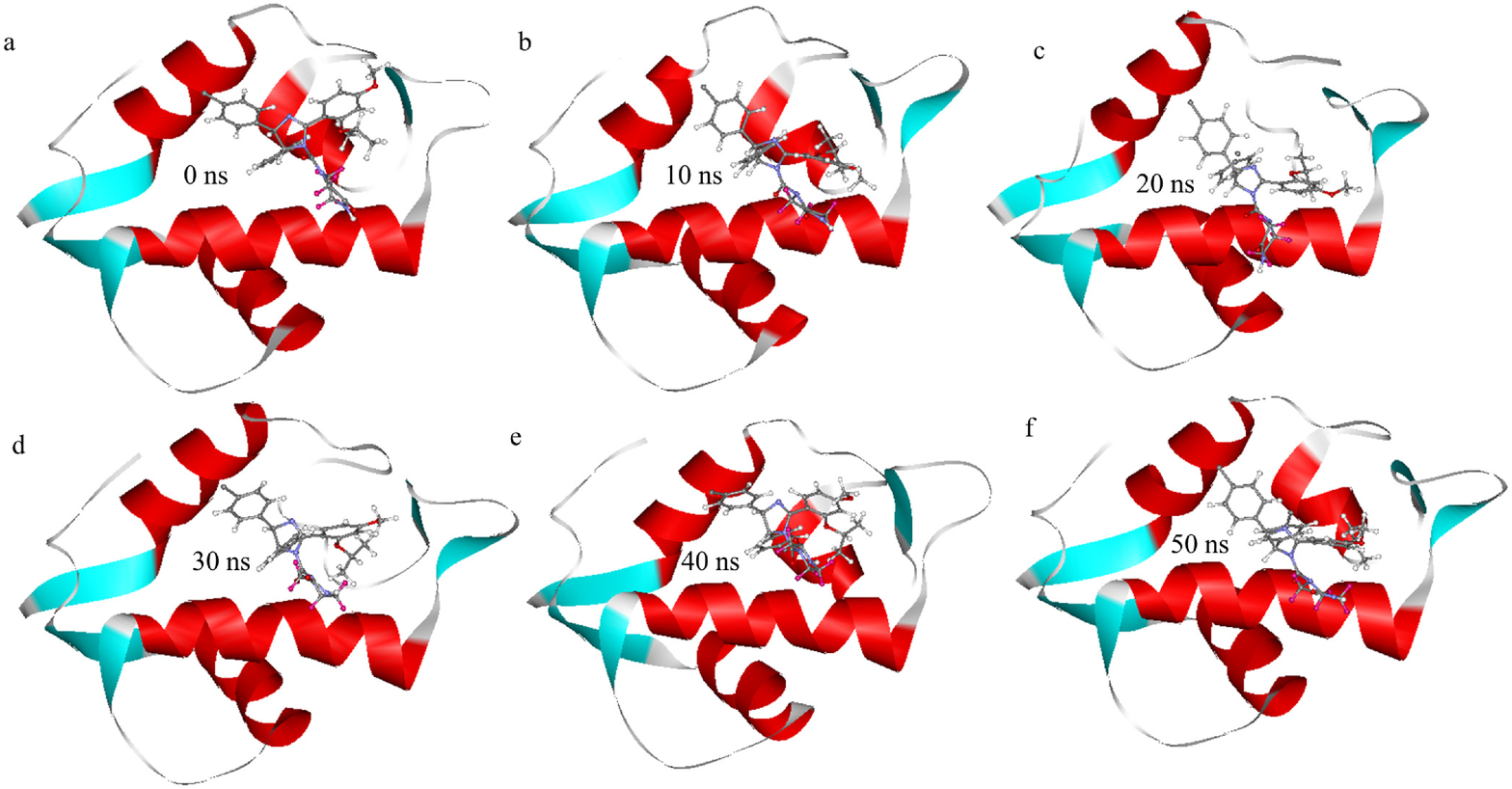}
\end{center}
\caption{Snapshots of MdmX-Nutlin3 complex at 0(a), 10(b), 20(c), 30(d), 40(e) and 50(f) ns in MD simulation. Nutlin3 molecule can bind to the p53-binding site of MdmX but highly flexible.}\label{fig4}
\end{figure*}

In order to investigate such structural differences, we compare the amino acid sequences and the tertiary structures of Mdm2/MdmX\textquoteright s p53-binding domain, as shown in Fig.~\ref{fig1}. It is clear that the sequence identity degree between N-termini of Mdm2 and MdmX is $53.9\%$ and the differences between them can be classified as three types with changes of polarity ($\sim$36.6\%), hydrophobicity ($\sim$19.5\%) and geometric structures of side chains ($\sim$39\%). There are four main regions in their backbone that show relatively large difference (C$_\alpha$s\textquoteright ~distances larger than 0.7 \AA) among proteins, which are residues Ser40 to Met50, Thr63 to Gln71, Ser78 to Phe86 and Lys94 to Arg105 in Mdm2 as in Fig.~\ref{fig1}b and Fig.~\ref{fig1}c, although the superposition of tertiary structures of corresponding region between Mdm2 and MdmX shows high similarity. The segments of Ser40 to Met50 and Thr63 to Gln71 are loops in secondary structures which show large flexibility in water environment \cite{CMB-2-177}. Therefore, the structural difference between Mdm2 and MdmX caused by sequence
differences is on two helices, H3 (Mdm2; residues 78-84 ) and H4 (Mdm2; residues 94-106), within which, H4 is the key element of p53 binding sites of Mdm2 and MdmX.

Note here that, Mdm2 and MdmX's $\alpha$ helix H1s show obvious difference. From crystal structures of p53-Mdm2, H1 of Mdm2 covers the p53-binding site of Mdm2, which indicates the potential function of H1 in the binding of two proteins. However, H1 of MdmX is absent in the crystal structures and we can not identify its exact position. We will discuss the roles of helix H1 further in the following Sec.~\ref{sect-3-3}.

\subsection{Binding patterns of p53 to Mdm2 and MdmX\label{sect-3-2}}
We also should keep in mind that structures and functions of a bio-molecule are bridged by dynamical process occuring from micro- to macro-scales. In order to clarify why the above-mentioned structural difference results in different binding characteristics between p53-Mdm2/p53-MdmX and between inhibitor-Mdm2/inhibitor-MdmX, we first compare binding pattern of p53 to Mdm2 and MdmX with MD simulation. It is known that both Mdm2 and MdmX proteins can bind to p53 with the same binding sites to negatively control the concentration of p53 in cells \cite{Nature-358-80,Nature-362-857}. To identify the binding patterns of p53 with Mdm2 and MdmX, we solvate the p53-Mdm2 and p53-MdmX complexes respectively to perform MD simulations. After 50 ns MD simulation, we obtained the stable complexes of p53\textquoteright s N-terminal helix and Mdm2/MamX\textquoteright s p53-binding sites, as shown in Fig.~\ref{fig2}a, b and c. The superposition of stable complex structures of p53-Mdm2 and p53-MdmX indicates that the binding patterns of p53 to Mdm2 and MdmX are same though there is some difference between Mdm2 and MdmX\textquoteright s sequences and structures.

We analyse structure of p53-Mdm2 and p53-MdmX complexes respectively and find that hydrophobic interactions and hydrogen bond interactions are important for the binding of p53 to Mdm2/MdmX (Same as in \cite{ACIE-50-2680,Science-274-948,Structure-21-2143,PNAS-100-164,JACS-130-6472,JACS-134-6855,Annu-Rev-Biochem-77-557}). The hydrophobic surfaces of the p53 binding sites on Mdm2 and MdmX are similar, as in Fig.~\ref{fig2}c. There are three hydrogen bonds formed between Phe19, Trp23 and Leu26 of p53 and Gln72/Gln71, Leu54/Met53 and Val93/Val92 of Mdm2/MdmX, as shown in Fig.~\ref{fig2}d. For p53-Mdm2 complex, such three hydrogen bonds formed by the hydrogen atom HE1 of residue Trp23 of p53 and the oxygen atom O of residue Leu54 of Mdm2 (HB1), the hydrogen atom HZ of residue Phe19 of p53 and the oxygen atom of OE1 of residue Gln72 of Mdm2 (HB2) and the hydrogen atom of H15 of residue Leu26 of p53 and the oxygen atom O of Val93 of Mdm2 (HB3), respectively. In these three hydrogen bonds, the most important and stable one is the hydrogen bond between Trp23 of p53 and Leu54/Met53 of Mdm2/MdmX. The same conclusion has been reported in previous studies \cite{NRC-3-102,MCR-23-1998}. The other two hydrogen bonds are not such stable since they locate at the margin of the complex models and expose to surrounding water molecules.

It is worthwhile to note that our MD simulation has realized the situation in which Phe19, Trp23, and Leu26 residues of p53 inserts into the hydrophobic cavity in Mdm2 to form hydrophobic interactions (Fig.~\ref{fig2}b). The total energy of three hydrogen bonds is $-12.68$ kcal/mol. This reproduces the experimental observations and theoretical analyses reported in various literatures, i.e.,  \cite{ACIE-50-2680,Science-274-948,NRC-3-102,CR-73-2587,Structure-21-2143,MCR-23-1998,PNAS-100-164,JACS-130-6472,JACS-134-6855,Annu-Rev-Biochem-77-557}.

\subsection{Binding of Nutlin3 with MdmX and Mdm2 proteins\label{sect-3-3}}

In this subsection, we analyse the binding characteristics of Nutlin3 with MdmX and Mdm2 proteins. As the same as in Sec.~\ref{sect-3-2}, we solvate the Nutlin3-Mdm2 and Nutlin3-MdmX complexes to perform MD simulations. To check the stability of this two systems, we calculate the root mean square deviation (RMSD) of the distance between Nutlin3 and Mdm2, as well as  Nutlin3 and MdmX, respectively (shown in Fig.~\ref{fig3}). It can be seen that Mdm2-Nutlin3 complex is fully stabilized after $\sim$5 ns and reaches a nearly stationary state. However, MamX-Nutlin3 complex seems loosely combine and can not become stable. This is consistent with the segmental mutagenesis experiments reported in \cite{JACS-136-18023}. The snapshots of MdmX-Nutlin3 complex conformations at various simulation moments in the trajectory are shown in Fig.~\ref{fig4}, which also shows that Nutlin3 can not bind to MdmX stably.

 We here note that, in Fig.~\ref{fig4}, there is a large amplitude motion of helix H3. In Fig.~\ref{fig4}c and d, H3 is not a standard helix structure. H3 locates in the bottom of the binding pocket of MdmX and effectively protects the interactions between p53/Nutlin3 and the binding pocket of MdmX. However, H3 itself is not so stable because it is exposed to the surrounding water molecules. Two hydrogen bonds (Asp79-Glu83 and Leu80-Leu84) which maintain the helix structure show fluctuation between forming and opening states. In this way, H3 exhibits large amplitude motions in our simulation.

\begin{figure}[htbp]
\setlength{\abovecaptionskip}{-5pt}
\begin{center}
\includegraphics[width=1.0\linewidth]{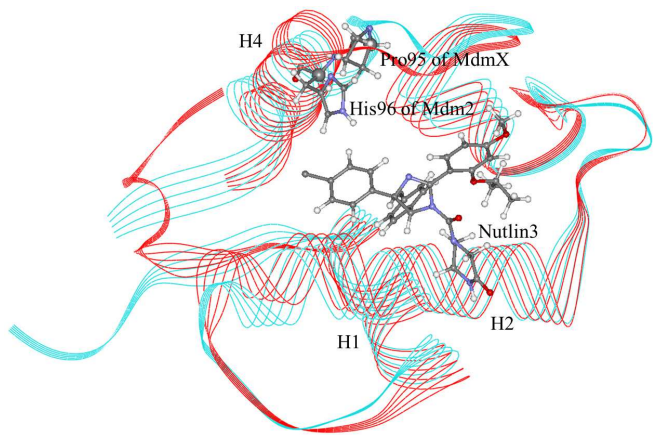}
\end{center}
\caption{Superposition of stable Mdm2 -Nutlin3 (red) and MdmX-Nutlin3 (blue) complex
structures. H2 and H4 helices of Mdm2 are close to each other relative to that of MdmX upon
binding with Nutlin3, that result in the different binding volume of Mdm2 and MdmX. The
His96 of Mdm2, corresponding to Pro95 of MdmX, covers the Nutlin3 in binding state,
which enhances the stability of inhibitor in the binding site of Mdm2.}\label{fig5}
\end{figure}

In order to understand above results deeper from atomistic level, we go further step to analyse the interactions between Nutlin3 with Mdm2 and MdmX and the dynamical evolution of their complex structures. Since the hydrogen bonding site of Nutlin3 is limit, Nutlin3 can not form stable hydrogen bond with Mdm2 or MdmX in our MD simulations. Therefore, we propose that the different binding behavior of Nutlin3 to Mdm2 and MdmX is caused by steric differences between Mdm2 and MdmX underlying the sequence differences between these two proteins. We superimpose the initial Mdm2-Nutlin3 and MdmX-Nutlin3 complexes (at the moment just after 30,000 steps energy minimization) of MD simulation to analyse their conformational difference as in Fig.~\ref{fig5}. Based on such analysis, we summarize the reasons why Nutlin3 can not bind to MdmX stably into the following three aspects.

\begin{figure}[htbp]
\setlength{\abovecaptionskip}{-15pt}
\begin{center}
\includegraphics[width=1.0\linewidth]{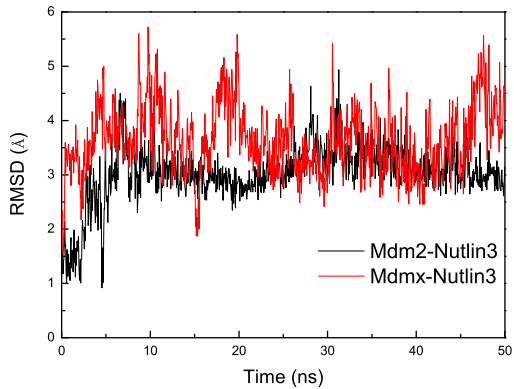}
\end{center}
\caption{RMSD of Ntulin3 in Mdm2-Nutlin3 without the extra $\alpha$ helix (black) and MdmX-Nutlin3 complex with the extra $\alpha$ helix (red) .}\label{fig6}
\end{figure}

First, in our simulation models, Mdm2 has an extra $\alpha$ helix (helix H1 in Fig.~\ref{fig1}) formed by residues Gln18 to Ala21 (Fig.~\ref{fig5}). This $\alpha$ helix effectively confine the movement of Nutlin3 molecule. However, the model of MdmX-Nutlin3 complex lack this $\alpha$ helical structure. We can not identify whether this $\alpha$ helix exists in the natural environment because there still lacks of the crystal structure of MdmX-Nutlin3 complex. To estimate the function of this extra $\alpha$ helix in the binding of Nutlin3 to Mdm2 or MdmX protein, we performed two MD simulations as, (1) Mdm2-Nutlin3 complex in which this extra $\alpha$ helix is deleted and (2) MdmX-Nutlin3 complex in which this extra $\alpha$ helix exists. The RMSD values of Ntulin3 in these two simulations are shown in Fig.~\ref{fig6}. It is clear that when lacking of the extra $\alpha$ helix, Mdm2-Nutlin3 complex becomes relatively instable compared with that when the extra $\alpha$ helix exists. Even in this case, such instable binding of Nutlin3 to Mdm2 is much more stable than that of Nutlin3 to MdmX without the extra $\alpha$ helix. When this helix exists, Nutlin3 can bind to MdmX stably. These results convincingly suggest that the extra $\alpha$ helix formed by Gln18 to Ala21 play a crucial role in stabilizing Nutlin3 molecule when its binding to Mdm2 or MdmX proteins.

Second, the volume sizes of Nutlin3-binding sites of Mdm2 and MdmX become different upon binding with Nutlin3. As discussed in Sec.~\ref{sect-3-1}, the hydrophobic surfaces of Mdm2 and MdmX are similar (shown in Fig.~\ref{fig2}c). However, seen from Fig.~\ref{fig5}, $\alpha$ helices H2 and H4 which form the p53 and inhibitor binding site of both Mdm2 and MdmX become different upon Nutlin3 binding. This difference results in the volume of inhibitor binding site of MdmX being larger than that of Mdm2 and thus Nutlin3 can not binding tightly to MdmX. This conformational difference between Mdm2 and MdmX\textquoteright s H2 and H4 arises from the sequence differences of two ends of H2 and H4 regions.

Third, when focusing on individual residues, we find that sequence difference at the following sites between Mdm2 and MdmX also contribute to the different binding behavior of Nutlin3 to Mdm2 and MdmX. The 96th site of Mdm2 is histidine and the corresponding site of MdmX is proline. The long side chain of His96 in Mdm2 can form stacking interaction with Nutlin3 and confine the movement of Nutlin3 molecule that further enhance the stability of Nutlin3 when binding to Mdm2 (Fig.~\ref{fig3}). However, Pro95 in MdmX can not form such interactions with Nutlin3.

At the end of this subsection, one may come to the following conclusions. There are various structural differences between Nutlin3-Mdm2 and Nutlin3-MdmX complex structures that result in the different binding characters of Nutlin3 to Mdm2 and MdmX. Such sequence-based structural differences are in three $\alpha$ helices (H1, H2 and H4) and one amino-acid site (His96 of Mdm2 and Pro95 of MdmX). One should sufficiently consider these differences of Mdm2 and MdmX for designing dual inhibitors of Mdm2 and MdmX and enhancing the stability of inhibitors in its binding site of MdmX.

\subsection{Disruption of Mdm2-p53 interaction with small molecules\label{sect-3-4}}

\begin{figure}[htbp]
\setlength{\abovecaptionskip}{-5pt}
\begin{center}
\includegraphics[width=0.95\linewidth]{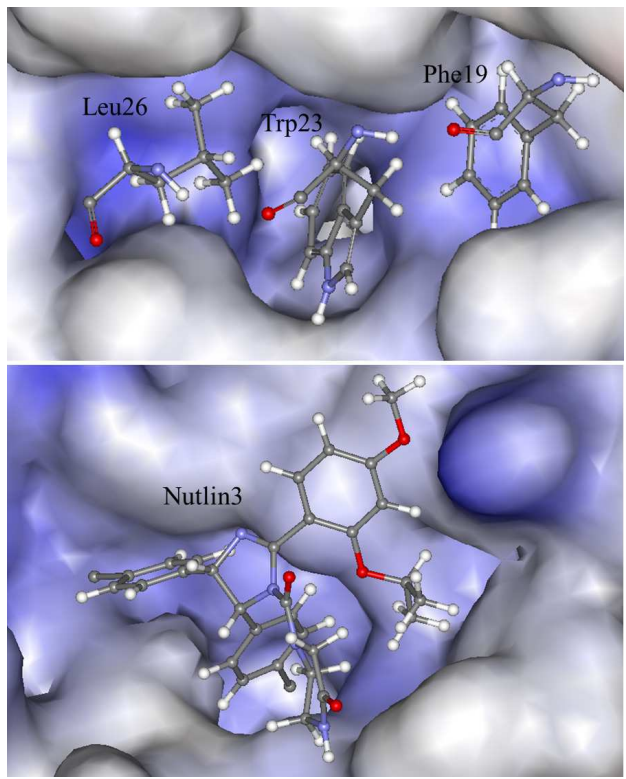}
\end{center}
\caption{p53 and Nutlin3 binding to the extended hydrophobic cleft on the N-terminus of Mdm2}\label{fig7}
\end{figure}

From the static binding point of view, small molecules like Nutlins can used to displace recombinant p53 protein from its complex with Mdm2 as shown in Fig.~~\ref{fig7}. However, it is still unclear how such small molecules dynamically to realize the disruption of Mdm2-p53 interaction and replacement of p53 from Mdm2-p53 complex {\it in vivo}, because, {initially}, inhibitors are far way from Mdm2-p53 complex in a water environment. This is an important issue to understand the action principle of inhibitors.

In general, such process of disruption and replacement is dynamical and is affected by the complex biological environment. In order to investigate whether Nutlin3 can disturb the interaction between p53 and Mdm2 and competes with p53 for binding to the extended hydrophobic-cleft on the N-terminus of Mdm2, we put a Nutlin3 molecule far away from the p53-Mdm2 complex and run the simulation for 180 ns. The whole system is in a water environment with temperature 310K.

\begin{figure*}[htbp]
\setlength{\abovecaptionskip}{-5pt}
\begin{center}
\includegraphics[width=0.85\linewidth]{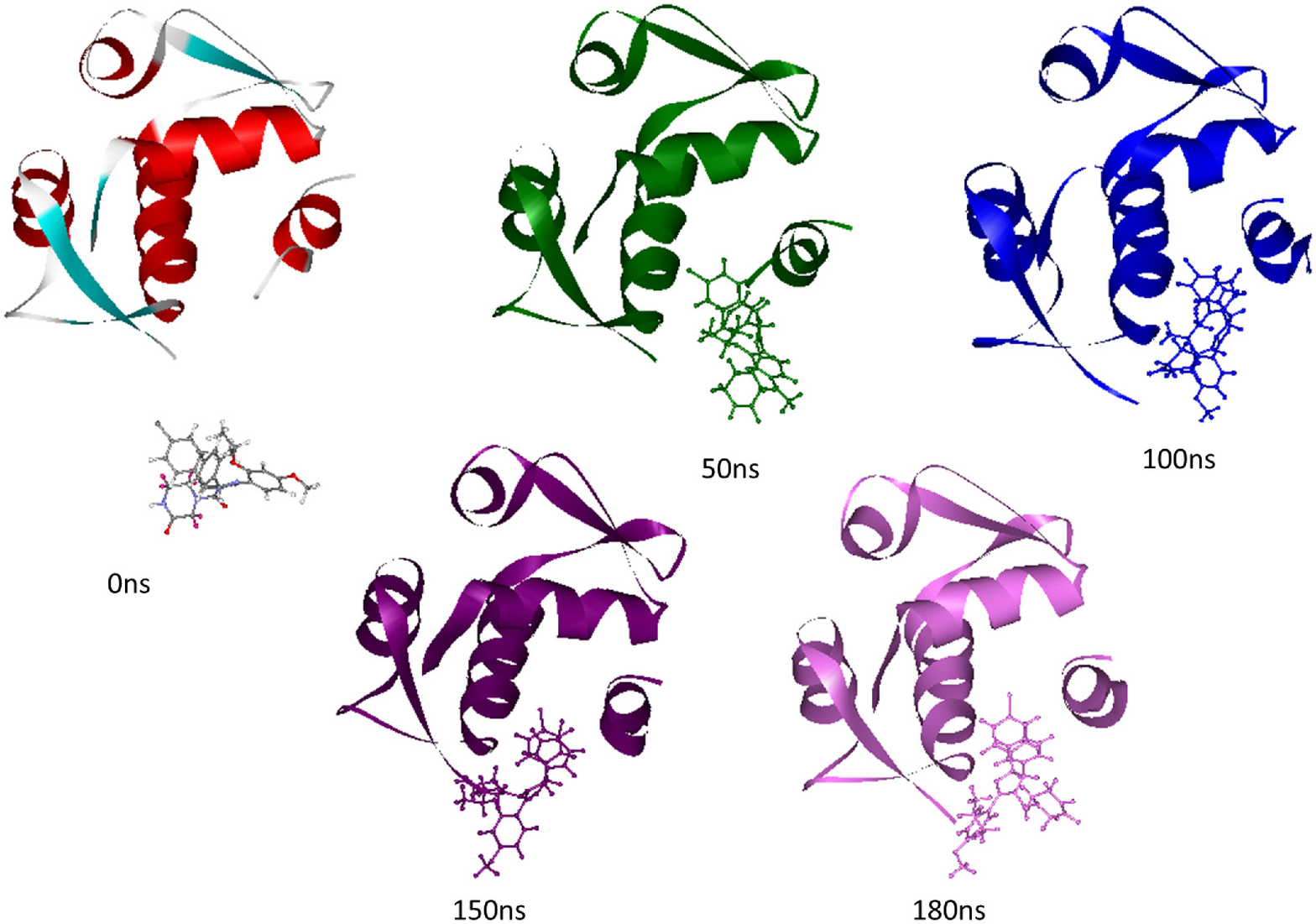}
\end{center}
\caption{Dynamical snapshots of Nutlin3-Mdm2-p53 complex at various moment of MD simulation}\label{fig8}
\end{figure*}

Five dynamical snapshots of the system are shown in Fig.~\ref{fig8}. Initially at 0 ns, p53 and Mdm2 bind stably, and Nutlin3 is far away from the complex. As mentioned above, in this initial case, there are three hydrogen bonds HB1, HB2 and HB3. Within in those three hydrogen bonds, HB1 is the most stable one. The interactions between p53 and Mdm2 in this initial snapshot is the same with the aforementioned simulations. At 50 ns, the interactions between p53 and Mdm2 are disturbed slightly, and the Nutlin3 moves closer to the binding site of Mdm2. HB1 is still stable, however, the other two hydrogen bonds (HB2 and HB3) has become rather weak and then even break down.  Till 100 ns, even HB1 also become weak.  At 150 ns through 180 ns, HB1 completely breaks down and p53 moves further away from the binding cavity, compared with the snapshot at 100 ns.

Above results intuitively and dynamically reveals that Nutlin3 can destroy the interactions between p53 and Mdm2,  go into the binding site of Mdm2 and finally form complex with Mdm2 in expectation (we hope so if the MD simulation ran long enough). This competition mechanism is the most important basis for understanding the inhibiting mechanism of small molecular inhibitors and cancer therapeutic strategy. Meaningfully, this microscopic study may establish a linkage with the macroscopic study of p53 pathway with stochastic dynamics \cite{CC-12-394,Proteomics-13-2512,PLOS-10-1003991}, for example, providing the important parameters like reaction rate constants and timescales.

\section{Concluding Remarks\label{sect-4}}

In this paper, we systematically investigated the structural characteristics and interaction properties of inhibitor-Mdm2 and -MdmX at atomistic level and compare it to the interaction of p53 with Mdm2 and MdmX to explore the molecular basis of inhibition. It is revealed that the structure of Mdm2 and MdmX are very similar except some amino acids in two $\alpha$ helices H3 and H4 of Mdm2 and/or MdmX. The formation of the complexes of p53-MdmX and p53-Mdm2 rely on the hydrophobic interactions and the hydrogen bond interactions.

We analyse the reason why Nutlin3 can combine to Mdm2 rather than MdmX. Now it is clear that there are three factors which influence the binding of Nutlin3 to Mdm2 and/or MdmX. First, an extra $\alpha$ helix at the N-terminus of Mdm2 and/or MdmX is important in the tightly binding of Nutlin3 to Mdm2 and MdmX. Second, in the dynamical docking process of Nutlin3 to Mdm2/MdmX, there are conformational changes in two $\alpha$ helices H2 and H4 which results in the different volume of inhibitor-binding site of Mdm2 and MdmX. Third, His96 in Mdm2 has an important role in the binding of Nutlin3 to Mdm2.

From dynamical point of view, we have proven that Nutlin3 can combine to Mdm2 and effectively inhibit the p53-Mdm2 interaction. Restoration the activity of p53 by inhibiting the interaction of p53-Mdm2 and/or MdmX is a promising and feasible method for cancer therapy, although its true therapeutic potential is yet to be elucidated.

It should be noted that the studies reported in this paper is still in a very initial stage. The detailed information needs to be verified through structural and dynamical investigation. In order to get a sufficient understanding of the inhibiting mechanisms, one has to combine methods, such as molecular docking method, MD simulation, free energy analysis and even stochastic network dynamics \cite{Yan-1}. Molecular docking method is expected to provide a reasonable initial structure of inhibitor molecules for MD simulation. Consider that bio-molecules are generally in a most heterogenous, extremely dynamical, far from equilibrium and complex fluctuating environment, stochastic network dynamics is needed to study p53 pathway \cite{Yan-2}. It is no doubt that such study is a challenging project. Let us leave it for further efforts.

\section*{Acknowledgements}
This work is supported by the National Nature Science Foundation of China (Grant Nos. 11675018, 10975019, 11605038, 11545014), Beijing Natural Science Foundation (Grant No. 1172008), the Foundation of the Ministry of Personnel of China for Returned Scholars (Grant No. MOP2006138) and the Fundamental Research Funds for the Central Universities  (Grant No. 2015KJJCB01)


\begin{thebibliography}{999}

\bibitem{ACIE-50-2680} G.~M.~Popowicz, A.~Domling, T.~A.~Holak, Angew.~Chem.~Int.~Ed.~{\bf 50}, 2680(2011).

\bibitem{JBC-286-23725} X. Wang, J. Wang, X. Jiang, J. Bio. Chem {\bf 286}, 23725(2011).

\bibitem{Science-303-844} L. T. Vassilev, B. T. Vu, B. Graves, et al., Science {\bf 303}, 844(2004).

\bibitem{Science-274-948} P. H. Kussie, S. Gorina, V. Marechal, et al., Science {\bf 274}, 948(1996).

\bibitem{NRC-3-102} P. Chene, Nat. Rev. Cancer {\bf 3}, 102(2003).

\bibitem{NRDD-3-301} M. R. Arkin, J. A. Wells, Nat. Rev. Drug. Discov. {\bf 3}, 301(2004).

\bibitem{JMM-83-955} D. C. Fry, L. T. Vassilev, J. Mol. Med. {\bf 83}, 955(2005).

\bibitem{Structure-21-2143} M. Bista, S. Wolf, K. Khoury, Structure {\bf 21}, 2143(2013).

\bibitem{NRDD-13-217} K. K. Hoe, C. S. Verma, D. P. Lane, Nat. Rev. Drug. Discov. {\bf 13}, 217(2014).

\bibitem{PBMB-117-250} T. Saha, R. K. Kar, G. Sa, Prog. Biophys. Mol. Biol. {\bf 117}, 250(2015).

\bibitem{CR-73-2587} S. Tovar, B. Graves, K. Packman, et al., Cancer Res. {\bf 73}, 2587(2013)

\bibitem{JACS-136-18023} L. Y. Qin, F. Yang, C. Zhou, et al., J. Am. Chem. Soc. {\bf 136}, 18023(2014)

\bibitem{MCR-1-1001}U. M. Moll, O. Petrenko,  Mol. Cancer Res. {\bf 1}, 1001 (2003).

\bibitem{EMBO-R-2-1029} R. Stad, N. A. Little, D. P. Xirodimas, et. al., EMBO Rep. {\bf 2}, 1029(2001).

\bibitem{ARPT-49-223} S. Shangary, S. Wang, Annu. Rev. Pharmacol. Toxicol. {\bf 49}, 223(2009).

\bibitem{JPCB-110-26365} M. D. M. AbdulHameed, A. Hamza, C. -G. Zhan, Annu. J. Phys. Chem. B {\bf 110}, 26365(2006).

\bibitem{CC-7-2441} G. M. popowicz, A. Czarna, T. A. Holak, Cell Cycle {\bf 7}, 2441(2008).

\bibitem{ACSD-69-1358} M. E. M. Noble, B. Anil, C. Riedinger, et al., Acta Crystallogr. Sect. D {\bf 69}, 1358(2013).


\bibitem{JCP-79-926} W. J. Jorgensen, J. Chandrasekhar, J. D. Madura, et al., J. Chem. Phys. {\bf 79}, 926(1983).

\bibitem{JMG-14-33} W. Humphrey, A. Dalke, K. Schulten, J. Mol. Graphics {\bf 14}, 33(1996).

\bibitem{JCC-26-1781} J. C. Phillips, R. Braun, W. Wang, et al., J. Comput. Chem {\bf 26}, 1781(2005).

\bibitem{JPC-102-3586} A. D. MacKerell, D. Bashford, M. Bashford, et al., J. Phys. Chem. {\bf 102}, 3586(1998).

\bibitem{PNAS-105-3933} S. Shangary, D. Qin, D. McEachern, et al., Proc. Natl. Acad. Sci. USA {\bf 105}, 3933(2008).

\bibitem{JBC-281-33030} B. Hu, D. M. Gilkes, B. Farooqi, et al., J. Biol. Chem. {\bf 281}, 33030(2006).

\bibitem{Nature-444-61} N. A. Laurie, S. L. Donovan, C. S. Shih,  et al., Nature {\bf 444}, 61(2006).

\bibitem{CMB-2-177} V. Hariharan and W. O. Hancock, Cell. Mol. Bioeng. {\bf 2}, 177 (2009).

\bibitem{Nature-358-80} J. D. Oliner, K. W. Kinzler, P. S. Meltzer, D. L. George, B. Vogelstein, Nature {\bf 358}, 80(1992).

\bibitem{Nature-362-857} J. D. Oliner, et al., Nature {\bf 362}, 857(1993).

\bibitem{MCR-23-1998} S. Md. A. Rauf, H. T. C. A. D. Carpio, A. Miyamoto, Med. Chem. Res. {\bf 23}, 1998(2014).

\bibitem{PNAS-100-164} M. A. McCoy, J.J. Gesell, M.M. Senior, et al., Proc. Natl. Acad. Sci. USA {\bf 100}, 164(2003).

\bibitem{JACS-130-6472} S. A. Showalter, L. Bruschweiler-Li, E. Johnson, et al., J. Am. Chem. Soc. {\bf 130}, 6472(2008).

\bibitem{JACS-134-6855} C. Y. Zhan, K. Varney, W. Y. Yuan, et al., J. Am. Chem. Soc. {\bf 134}, 6855(2012).

\bibitem{Annu-Rev-Biochem-77-557} A. C. Joerger, A. R. Fersht, Annu. Rev. Biochem. {\bf 77}, 557(2008).

\bibitem{CC-12-394} K. M. ElSawy, C. S. Verma, T. L. Joseph, et al., Cell. Cycle 12, 394(2013).

\bibitem{Proteomics-13-2512} L. Hernychova, P. Man, C. Verma, et al., Proteomics {\bf 13}, 2512(2013).

\bibitem{PLOS-10-1003991} K. Puszynski, A. Gandolfi and A. d’Onofrio, PLOS Computat. Biol. {\bf 10}, e1003991(2014).

\bibitem{Yan-1} S. X. Liu, Y. Z. Geng and S. W. Yan, to be published in Prog. Biochem. Biophys.

\bibitem{Yan-2} B. Liu, S. W. Yan and X. F. Gao, PloS ONE, {\bf 6}, e22487(2011) .

\end{thebibliography}
\end{document}